\documentclass[pra,%
 aps,twocolumn,%
 amsmath,amssymb,
 superscriptaddress,
 reprint,%
]{revtex4-2}
\usepackage{booktabs}
\usepackage{graphicx}
\usepackage{dcolumn}
\usepackage{bm}
\usepackage{graphicx} 
\usepackage{multirow} 
\usepackage{etoolbox}
\usepackage{braket}
\usepackage{hyperref}
\usepackage{algorithm}
\usepackage{algpseudocode}
\usepackage{setspace}
\usepackage{float}
\usepackage{placeins}
\usepackage{multirow}
\usepackage{booktabs}
\usepackage{nicematrix}
\usepackage{todonotes}
\usepackage{physics}
\usepackage{amsthm} 
\usepackage{ bbold }
\usepackage[normalem]{ulem}
\definecolor{forestgreen(web)}{rgb}{0.13, 0.55, 0.13}
\usepackage{cleveref}
\usepackage{lipsum}

\usepackage{algorithm}
\usepackage{algpseudocode}
\usepackage{amsmath}

\algnewcommand{\algorithmicor}{\textbf{ or }}
\algnewcommand{\continue}{\textbf{continue}}
\algnewcommand{\OR}{\algorithmicor}

\makeatletter
\newcommand{\bigcomp}{%
  \DOTSB
  \mathop{\vphantom{\sum}\mathpalette\bigcomp@\relax}%
  \slimits@
}

\newcolumntype{C}[1]{>{\centering\arraybackslash}m{#1}}
\newcommand{\bigcomp@}[2]{%
  \begingroup\m@th
  \sbox\z@{$#1\sum$}%
  \setlength{\unitlength}{0.9\dimexpr\ht\z@+\dp\z@}%
  \vcenter{\hbox{%
    \begin{picture}(1,1)
    \bigcomp@linethickness{#1}
    \put(0.5,0.5){\circle{1}}
    \end{picture}%
  }}%
  \endgroup
}
\newcommand{\bigcomp@linethickness}[1]{%
  \linethickness{%
      \ifx#1\displaystyle 2\fontdimen8\textfont\else
      \ifx#1\textstyle 1.65\fontdimen8\textfont\else
      \ifx#1\scriptstyle 1.65\fontdimen8\scriptfont\else
      1.65\fontdimen8\scriptscriptfont\fi\fi\fi 3
  }%
}

\makeatother
\begin{document}

\preprint{AIP/123-QED}

\title{Machine Learning-based Quantum Error Mitigation for Variational Algorithms}

\author{Nikita Korolev}%
 \email{n.korolev@rqc.ru}
 \affiliation{Russian Quantum Center, Moscow, Russian Federation}%



\author{Kirill Lakhmanskiy}%
\affiliation{Russian Quantum Center, Moscow, Russian Federation}%

\author{Daniil Rabinovich}%
 \email{D.rabinovich@rqc.ru}
\affiliation{Russian Quantum Center, Moscow, Russian Federation}%
\affiliation{Skolkovo Institute of Science and Technology, Moscow, Russian Federation}%
 \affiliation{Moscow Institute of Physics and Technology, Moscow, Russian Federation}
 
\date{\today}

\begin{abstract}
Machine Learning-based Quantum Error Mitigation (ML-QEM) has emerged as a promising approach for improving the performance of noisy quantum algorithms. However, existing ML-QEM methods often have restricted applicability to variational circuits and rely on inaccessible noiseless training data. In this work, we propose a practical ML-QEM protocol tailored to variational quantum algorithms, which generates training data by simulating (near-)Clifford circuits. This data is used for model selection and training, producing a mitigation model that can correct variational circuits with arbitrary parameters and transfer across different target Hamiltonians of similar structure. We benchmark the proposed method on the Variational Quantum Eigensolver (VQE) task for the Sherrington-Kirkpatrick Hamiltonian of up to $n=12$ qubits under various noise models, analyzing its effect on trainability and comparing its performance against standard Zero-Noise Extrapolation (ZNE). The results demonstrate consistent several-fold error suppression across all tested settings and superior performance over ZNE in the high-noise regime, providing evidence for the applicability of the proposed protocol to present-day NISQ processors.

\end{abstract}

\maketitle
\section{Introduction}
Present-day quantum computers are represented by Noisy Intermediate Scale Quantum (NISQ) devices \cite{preskill2018}. These devices are limited by their short coherence times, moderate system sizes, and limited fidelities of entangling operations \cite{weidenfeller2022, ratcliffe2018, akopyan2022, hegde2022, mills2022}. Performing fault-tolerant quantum computing would require implementation of Quantum Error Correction (QEC) \cite{knill2000theory} protocols, which are not achievable with the current scope of hardware \cite{chiaverini2004realization}. Faced with these limitations, several alternatives have been proposed for early implementations of quantum computations.

Variational Quantum Algorithms (VQAs) \cite{cerezo2021, bharti2022} are hybrid quantum-classical algorithms designed to find approximate solutions to complex problems using current NISQ devices.  They combine classical optimization with quantum computation, leveraging parameterized quantum circuits—known as ansatz—to explore solution spaces.  The quantum computer evaluates a cost function (e.g., expectation value of a Hamiltonian), while a classical optimizer iteratively adjusts the circuit parameters to minimize this cost, aiming for an approximate solution.

Even though VQAs are tailored to operate on noisy
hardware, they still suffer from infidelities of quantum
gates \cite{huang2023}. Despite recent advancements in quantum gate precision, hardware errors still detrimentally affect algorithmic performance \cite{proctor2022measuring}. Utilizing QEC~\cite{devitt2013} requires lower error thresholds and a number of qubits far beyond modern capabilities. Thus, other approaches, such as Quantum Error Mitigation (QEM) \cite{cai2023}, have been developed to assist in the early implementations of quantum algorithms. QEM, unlike QEC, does not aim to physically suppress gate errors but instead attempts to recover noiseless expectation values through measurement post-processing.

Existing QEM techniques can be divided into noise-aware \cite{Temme_2017, gupta2023probabilisticerrorcancellationdynamic} and noise-agnostic methods \cite{Giurgica_Tiron_2020, czarnik2021}. Noise-aware approaches require knowledge about the noise in the system, for example, to approximate its mathematical inverse and suppress the noise effect on the final result. However, acquiring this noise model might appear quite challenging in practice as it requires full noise tomography \cite{henao2023adaptive}. Noise-agnostic approaches, on the contrary, do not require such knowledge. Nonetheless, they often come at the cost of reduced mitigation accuracy due to its ``black box" nature.

Recently proposed data-driven quantum error mitigation methods \cite{liao2024, strikis2021learning, kim2020quantum, muqeet2024machine} can be viewed as an intermediate ``gray box" approach between these paradigms. While requiring no prior knowledge of the underlying noise, machine learning models are trained on collected data to suppress errors by approximating the inverse of the noisy channel. One of the key challenges in applying Machine Learning-based Quantum Error Mitigation (ML-QEM) techniques is gathering the training dataset. Indeed, while noisy expectation values can be directly obtained on a quantum device, obtaining noiseless expectation values becomes non-trivial. Numerous proposals have been made to tackle this issue: using classical simulations, results of other QEM methods \cite{liao2024}, echo-evolution \cite{babukhin2024echo} and efficiently simulated classically (near-)Clifford circuits \cite{czarnik2021}. The latter approach is of great interest due to its high practicality. However, the scope of existing works limits their consideration to circuits with fixed parameters, which might have limited applicability to variational computing. 

In this work, we develop and systematically benchmark a practical framework for machine learning–based quantum error mitigation in variational quantum algorithms. In particular, we provide an extensive comparison of different regression models trained on data generated from Clifford and near-Clifford circuits. The method is shown  to be capable of constructing effective mitigation maps across a range of noise models and noise strengths considered. We further evaluate several regression models, including Ridge regression \cite{mcdonald2009ridge} and XGBoost \cite{chen2016xgboost}, for reconstructing noiseless Hamiltonian expectation values from noisy circuits and compare their performance against standard Zero-Noise Extrapolation (ZNE). This analysis allows us to identify the most suitable model class for this task and to investigate the trade-off between model complexity and robustness. Finally, we study different mitigation regimes— post-optimization correction and in-loop mitigation—as well as the transferability of trained models across ansatze with similar entangling structures.

The paper is organized as follows. Section~\ref{prelim} introduces the theoretical background on variational quantum algorithms, the considered noise models, and the formulation of machine learning-based quantum error mitigation. Section~\ref{main} presents the proposed ML-QEM protocol, including dataset generation, model selection and training, and its integration into the VQA workflow, followed by numerical benchmarking and comparison with ZNE. Finally, Section~\ref{conclusion} summarizes the main results and discusses the implications and limitations of the proposed approach.

\section{\label{sec:prelim}Preliminaries}
\label{prelim}
\subsection{Variational quantum algorithms}

One of the most widely studied types of VQAs is Variational Quantum Eigensolver (VQE) \cite{tilly2022}, which searches for the ground state energy of a quantum system — which finds applications in quantum chemistry simulations \cite{li2019variational, feniou2023overlap} and condensed matter physics \cite{ma2020quantum, sun2023efficient}. The approach is inspired by the variational principle, which ensures that
\begin{equation}
   E_0 \leq \dfrac{\bra{\psi}H\ket{\psi}}{\langle\psi|\psi\rangle},
\end{equation}
where $H$ is the objective Hamiltonian, $\ket{\psi}$ is a trial state vector, and $E_0$ is the ground state energy  of the Hamiltonian $H$. Thus, the objective of VQE is to find the trial quantum state that minimizes the Hamiltonian expectation value. In other words, one aims to approximately find the eigenvector $\ket{\psi}$ of a Hamiltonian $H$ with the lowest eigenenergy $E_0$.

The trial quantum state is prepared on a quantum computer using a parametrized quantum circuit $U(\bm{\theta})$ with $N$ qubits, where $\bm{\theta}$ is a vector of parameters $\theta_j$, each taking values, for example, from $(-\pi,\pi]$. In this approach, qubits are initialized in an easy to prepare quantum state, e.g. $\ket{0}^{\otimes N} = |\mathbf{0}\rangle$. Then the VQE optimization problem can be written as
\begin{equation}\label{eq:cost_func}
    E_{\text{VQE}} = \min_{\bm{\theta}} \bra{\mathbf{0}}U^\dagger(\bm{\theta})HU(\bm{\theta})\ket{\mathbf{0}} = \min_{\bm{\theta}}C(\bm{\theta}),
\end{equation}
where $C(\bm{\theta})$ is called a cost function. 

A parametrized quantum circuit consists of two classes of gates: fixed gates, such as CNOTs, and parametrized gates, which are usually represented by single-qubit rotations $R_X(\theta), R_Y(\theta)$ and $R_Z(\theta)$. The arrangement of these gates, i.e.~the manner in which the quantum parametrized circuit is composed from them, is called an \textit{ansatz}.  

A problem Hamiltonian can typically be presented as a weighted sum of Pauli strings,
\begin{equation}\label{eq:ham}
    H = \sum\limits_{\alpha=1}^\mathcal{P}h_\alpha P_\alpha,
\end{equation}
where $P_\alpha \in \{\mathbb{1}, X, Y, Z\}^{\otimes n}$ is a Pauli string with $X, Y, Z$ being corresponding Pauli operators, and $n$ is the number of qubits. Here $h_{\alpha}$  are corresponding weights and $\mathcal{P}$ is the number of Pauli strings in the Hamiltonian or a so-called Hamiltonian's cardinality. Taking this into account, the optimization task \eqref{eq:cost_func} can be rewritten as 
\begin{equation}
     E_{\text{VQE}} = \min_{\bm{\theta}} \sum\limits_{\alpha=1}^\mathcal{P}h_\alpha\bra{\mathbf{0}}U^\dagger(\bm{\theta})P_\alpha U(\bm{\theta})\ket{\mathbf{0}}.
\end{equation}
The hybrid nature of VQE becomes transparent in this expression. The quantum computer executes the parametrized circuit $U(\bm{\theta})$ and obtains the trial state $|\psi(\bm{\theta})\rangle = U(\bm{\theta})|\bm{0}\rangle$ and then each Pauli string term  $\langle P_\alpha\rangle(\bm{\theta}) = \langle\psi(\bm{\theta})|P_\alpha|\psi(\bm{\theta})\rangle = \bra{\mathbf{0}}U^\dagger(\bm{\theta})P_\alpha U(\bm{\theta})\ket{\mathbf{0}}$ from the resulting sum \eqref{eq:ham} is measured, which might require running the circuit several times. The resulting energy expectation value is computed on a classical computer as $E(\bm{\theta}) = \sum\limits_{\alpha=1}^\mathcal{P}\langle P_\alpha\rangle(\bm{\theta})$ which is later used in a classical optimizer~\cite{bonet2023}.

\subsection{\label{noise}Noise models}
As VQA performance is still heavily affected by the gate errors~\cite{huang2023}, the impact of quantum noise on such algorithms should be taken into consideration and studied. Due to the recent progress achieved in terms of fidelity of single-qubit gates across all platforms~\cite{tannu2018}, in this work we consider only two-qubit gate errors, modeled as quantum noisy channels $\rho \to \mathcal{E}(\rho)$ applied after every ideal two-qubit operation, thereby causing the resulting noisy expectation value to deviate from its noiseless value. Three major noise models are considered: 
\begin{enumerate}
    \item Depolarizing noise \cite{dur2005}, one of the most commonly used noise models in theoretical studies,  with the  quantum channel
    \begin{equation}\label{depol}
        \mathcal{E}(\rho)_{\text{depol}} = \dfrac{p}{d}\mathbb{1}_d + (1-p)\rho,
    \end{equation}
    where $\mathbb{1}_d$ is a $d$-dimensional identity. For modeling two-qubit gate errors, we assume the noise affects only the specific qubits considered, i.e.~$d=4$.
    \item Pauli noise \cite{flammia2020}, a more general, asymmetric version of the former model, with a single-qubit quantum channel of the form 
    \begin{align}
        \mathcal{E}(\rho)_{\text{Pauli}} = (1 - p_x - p_y - p_z)\rho &+ p_xX\rho X \nonumber\\+ p_yY\rho Y &+ p_zZ\rho Z,
    \end{align}
    where $p_x, p_y,p_z$ are the probabilities of applying the corresponding Pauli operators. In the scope of this work, we parametrize the single-qubit Pauli channel with a single strength parameter $p$ such as $p = p_x+p_y+pz$ and $p_x = p/3, p_y = 2p/9, p_z = 4p/9,$ thereby fixing the asymmetry of the channel.
    Two-qubit Pauli noise channel with strength $p$ is modeled as a tensor product of two single-qubit noisy channels $\mathcal{E}\otimes\mathcal{E}$ with strengths $p/2$ each. 
    \item Composite noise model~\cite{Georgopoulos_2021} consisting of three noise channels: depolarization \eqref{depol}, amplitude damping ~\cite{khatri2020} and phase damping~\cite{Czerwi_ski_2016} channels. A single-qubit amplitude damping channel with damping rate $\gamma$ is described as
    \begin{equation}\label{amp}
        \mathcal{E}_{\text{amp}}(\rho) = E_0\rho E_0 + E_1\rho E_1, 
    \end{equation}
    where 
    \begin{equation}
        E_0 = \begin{bmatrix}
1 & 0 \\
0 & \sqrt{1 - \gamma}
\end{bmatrix}
, \,\,\, E_1 = \begin{bmatrix}
0 & \sqrt{\gamma}\\
0 & 0
\end{bmatrix}.
    \end{equation}
    
    A single-qubit phase damping channel with a dephasing rate $\lambda$ is given by
    \begin{equation}\label{phas}
    \mathcal{E}_{\text{ph}}(\rho) = K_0\rho K_0 + K_1 \rho K_1,
    \end{equation}
    \begin{equation}
        K_0 = \begin{bmatrix}
1 & 0 \\
0 & \sqrt{1 - \lambda}
\end{bmatrix}
, \,\,\, K_1 = \begin{bmatrix}
0 & 0\\
0 & \sqrt{\lambda}
\end{bmatrix}.
    \end{equation}
To model realistic two-qubit gate imperfections, we consider a channel obtained by sequentially applying amplitude damping~\eqref{amp}, phase damping|\eqref{phas}, and depolarizing noise~\eqref{depol} after each ideal two-qubit gate. The resulting channel is defined as
\begin{equation}
    \mathcal{E}_{\text{tot}} = \mathcal{E}_{\text{depol}}\circ\left(\mathcal{E_{\text{ph}}}\otimes\mathcal{E_{\text{ph}}} \right)\circ\left(\mathcal{E_{\text{amp}}}\otimes\mathcal{E_{\text{amp}}} \right).
\end{equation}
In simulations, the amplitude- and phase-damping strengths were chosen as
\begin{equation}
    \gamma = \lambda = \dfrac{p}{2},
\end{equation}
such that the total noise strength is parametrized by a single effective two-qubit error rate $p$.
\end{enumerate}
These noise models are widely used to model the effect of noise on the algorithmic performance \cite{Bharti_2022}. In case of VQA, such models can affect the algorithm, degrading the quality of the found solution in two interconnected ways. First, the state prepared by the finally trained circuit degrades under the noise, which reduces the quality of the solution. Second, the presence of noise can affect the optimization process itself, leading to different circuit optimal parameters \cite{cai2023}.
\subsection{Quantum error mitigation as a machine learning task}

All existing QEM methods can be broadly divided into noise-aware and noise-agnostic approaches. A representative of the former is probabilistic error cancellation (PEC) \cite{van2023}, which relies on an accurate noise model $\Lambda$ to construct an approximate inverse $\Lambda^{-1}$ and recover noiseless expectation values.~However, this requires precise noise characterization—often challenging in practice—and incurs a significant sampling overhead due to increased estimator variance \cite{xiong2020sampling}. In contrast, noise-agnostic methods such as ZNE \cite{majumdar2023} avoid explicit noise modeling by evaluating expectation values at amplified noise levels and extrapolating to the zero-noise limit. While attractive, ZNE is constrained by the necessity to controllably scale noise and by limitations inherent to its black-box nature. One of the subdomains of noise-agnostic QEM methods are data-driven machine learning approaches \cite{liao2024, kim2020}. Having no prior access to the noise in the system, an ML model is trained to construct its approximate inverse. Thus, initially being noise-agnostic, this approach is no longer a fully ``black box".

Machine learning quantum error mitigation task can be formulated as constructing a parametrized map $f_\phi$ from noisy Pauli strings expectation values $\langle P_\alpha \rangle^\text{noisy}$ to their noiseless counterparts $\langle P_\alpha \rangle^\text{ideal}$. Formally, for a Hamiltonian expressed as a weighted sum of Pauli strings $H = \sum\limits_{\alpha=1}^\mathcal{P}h_\alpha P_\alpha$ it is required to construct the map
\begin{equation}\label{eq:pauli_mitig}
    f_\phi: [-1,1]^\mathcal{P} \to [-1,1]^\mathcal{P}, \,\,\, \mathbf{P}^{\text{noisy}} \to \hat{\mathbf{P}},
\end{equation}
where $\mathbf{P} = \left(\langle P_1\rangle, \langle P_2\rangle, \dotso, \langle P_\mathcal{P}\rangle\right)^T$, such that the final mitigated energy
\begin{equation}
    \hat{E} = \sum\limits_{\alpha = 1}^\mathcal{P}h_\alpha\hat{\langle P_\alpha\rangle} = \mathbf{h}\cdot\hat{\mathbf{P}}
\end{equation}
would approximate noiseless energy $E^{\text{ideal}}$. 

The corresponding map is obtained via training a machine learning model on a dataset $\left\{\mathbf{P}^{\text{noisy}}_i, \mathbf{P}^{\text{ideal}}_i\right\}_{i=1}^N$, where $N$ is the size of the dataset. The training is performed by minimizing the loss function
\begin{equation}
    L(\phi) = \frac{1}{N}\sum\limits_{i=1}^N||f_\phi\left(\mathbf{P}_i^{\text{noisy}}\right) - \mathbf{P}_i^{\text{ideal}}||.
\end{equation}
Alternatively, this task can be directly formulated as problem-specific Hamiltonian mitigation and constructing the map $f_\phi$ as
\begin{equation}\label{eq:ham_mitig}
      f_\phi: [-1, 1]^\mathcal{P} \to \mathbb{R}, \,\,\, \mathbf{P}^{\text{noisy}} \to \hat{E}.
\end{equation}
training ML models on the dataset  $\left\{\mathbf{P}_i^{\text{noisy}}, E^{\text{ideal}}_i \right\}_{i=1}^N$, where $E^{\text{ideal}}_i$ is the noiseless energy that corresponds to noisy Pauli strings $\mathbf{P}_i^{\text{noisy}}$. This approach proposes to construct the direct map from noisy Pauli strings to noiseless energy estimate $\hat{E}$. While this approach is computationally simpler, the method \eqref{eq:pauli_mitig} proposes a more versatile framework---once trained, it can be applied to a broad class of Hamiltonians composed of the same Pauli operators. 

Within the scope of this work, we focus on applying ML-QEM to VQE. In this setting, it is essential that the mitigation protocol remains efficient across the full range of variational parameters. Integrating ML-QEM into the VQE workflow raises a question~\cite{Wang_2024} if performing error mitigation during the optimization process influences the parameter updates and guides the algorithm toward more favorable solutions. In other words, should mitigation be applied as an integral part of the optimization loop or simply after the parameters have been optimized? This question, together with the comparison between \eqref{eq:pauli_mitig} and \eqref{eq:ham_mitig}, is investigated in the following section.

\section{Machine learning quantum error mitigation}
\label{main}
This section outlines the pipeline of the proposed protocol and presents results of numerical experiments and their discussion. We describe data acquisition for training the machine learning algorithm, followed by the selection of the best-performing model and its optimal hyperparameters, which are then integrated into the VQE framework, discuss various error mitigation regimes, compare their performance, and identify the most suitable approach. All quantum circuit simulations are conducted using the Qiskit library \cite{qiskit}. For machine learning, the XGBoost library \cite{chen2019package} is used specifically for the gradient-boosted tree model, while all other algorithms are implemented using scikit-learn \cite{pedregosa2011scikit}. The best models are subsequently tested against ZNE using different noise models, with randomly sampled Sherrington–Kirkpatrick Hamiltonians serving as the VQE target Hamiltonian. 
\subsection{\label{sec:mlqem_dataset_generation}Dataset generation}
An essential component of applying machine learning methods to quantum error mitigation, as in any supervised learning setting, is the construction of a training dataset $\left\{\mathbf{P}^{\text{noisy}}_i, \mathbf{P}^{\text{ideal}}_i\right\}_{i=1}^N$. For a given parametrized quantum circuit, the straightforward approach would be to sample $N$ sets of parameters and evaluate noiseless and noisy expectations of Pauli strings for each set. While noisy ones can be directly obtained via executing the parametrized circuit on the quantum computer, noiseless expectations can only be acquired from classical simulations. This method, however, is limited by the capabilities of classical computers, which are not powerful enough to simulate the number of qubits required to handle practical, real-world problems~\cite{chen201864}. 

A possible way to overcome this limitation is to restrict attention to quantum circuits composed solely of Clifford gates which, by the Gottesman–Knill theorem \cite{aaronson2004}, can be efficiently simulated classically. Circuits containing a small number of non-Clifford gates (e.g., single-qubit rotations) can also be efficiently simulated, although the computational cost grows exponentially with the number of such gates. These near-Clifford circuits are of particular interest in the context of machine learning-based quantum error mitigation, as they provide broader coverage of Hilbert space and thus could provide a more informative training dataset. In the following, we refer to circuits composed only of Clifford gates as Clifford circuits and to those containing a limited number of non-Clifford gates as near-Clifford circuits. We compare the resulting performance of ML-based error mitigation trained on datasets generated from both classes of circuits. 

Existing works that use Clifford-based datasets for training ML models in quantum error mitigation typically focus on fixed circuits, sampling (near-)Clifford circuits that are close to the target circuit in terms of observable values \cite{czarnik2021, lowe2021unified}. In contrast, here we  aim to train models capable of mitigating variational quantum circuits, i.e.~ansatz circuits with arbitrary parameters. Constructing the sample circuits for the dataset consists of replacing each parametrized operation in the considered ansatz with a randomly sampled Clifford gate, which results in a circuit composed of Clifford operations. To create near-Clifford circuits, we modify this approach by placing a layer of  random unitary gates sampled according to the Haar measure \cite{tang2022generating} at the beginning of the circuit---these would be the only non-Clifford gates in the circuit. However, as the number of qubits increases, simulating the generated near-Clifford circuits can also become intractable due to the high number of non-Clifford unitaries. To address this, these unitaries can be introduced probabilistically---applied to each qubit with a fixed probability $q$---thereby reducing the total number of non-Clifford gates in the circuit. Further, we analyze how this approach impacts the performance of ML-QEM.

\begin{figure}
    \centering
    \includegraphics[width=0.8\linewidth]{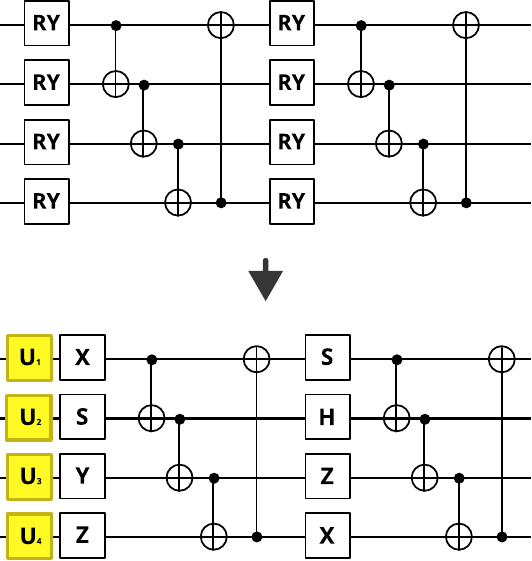}
    \caption{The dataset generation protocol scheme tailored to the \texttt{TwoLocal} ansatz. Yellow single-qubit gates $U_i$ represent random Haar unitaries.}
    \label{fig:dataset}
\end{figure}

In the scope of this work, both in the Clifford and near-Clifford scenarios, we tailored the consideration to a particular type of VQE ansatz with a ring topology of CNOT gates in the entangling layer. The circuit is built by alternating layers of CNOT gates with layers of single-qubit rotations. This ansatz is referred to as the \texttt{TwoLocal} ansatz~\cite{qiskit} and it is widely used in variational quantum algorithms due to its hardware-efficient structure and flexibility in generating highly entangled quantum states with relatively shallow circuits. The process for creating a sample circuit is illustrated in figure \ref{fig:dataset}. As this ansatz already uses Clifford entangling operations (CNOTs) the circuit generation requires replacing only single qubit rotations with Clifford operations. It is important to note that the resulting circuits are not unique to the specific variational ansatz used; tailoring the consideration to any other ansatz with a similar entangling layer topology but different single-qubit rotations would yield the same dataset. Overall, we generate circuits consisting of 4 layers applied to $n=12$ qubits. The target Hamiltonian for the VQE task is the widely studied Sherrington-Kirkpatrick model \cite{panchenko2012sherrington, sherrington1975solvable}
\begin{equation}\label{eq:sk}
    H = \dfrac{1}{\sqrt{n}}\sum\limits_{j> i}^nJ_{ij}Z_iZ_{j} + h\sum\limits_{i=1}^nX_i,
\end{equation}where $J_{ij} \sim  \mathcal{N} (0, 1)$, $Z$ and $X$ are the Pauli operators.
The transverse field strength $h$ is set to $1$. To apply the ML-QEM to this Hamiltonian corresponding Pauli strings $\langle Z_iZ_j\rangle$ and $\langle X_i\rangle$ require measuring. 
\vspace*{2mm}
In this work we consider two-qubit error rates of $\{0.01, 0.05, 0.1\}$ for all the noise models from Sec.~\ref{noise}. These values cover the spectrum of noise present in existing devices \cite{dahlhauser2021}.
After the generation, all the constructed circuits are executed on both noiseless and noisy simulators in Qiskit. All the considered noise models are simulated and the performance is compared. Each generated dataset consists of 1500 samples with each sample given by a pair $\left(\mathbf{P} _i^{\text{noisy}}, \mathbf{P} _i^{\text{ideal}}\right)$.

\subsection{\label{sec:results}Model selection \& training}

To ensure an unbiased evaluation of machine learning models, their hyperparameters must be adjusted. We perform this via grid search, i.e.~by training and evaluating models over a predefined set of hyperparameter combinations. These parameters (e.g.~the number of estimators in Random Forest or the learning rate in MLP and XGBoost) are then fixed prior to the model training and are not learned from the data.

Model performance is assessed using K-fold cross-validation to reduce bias from dataset splitting. The dataset is partitioned into $K$ folds; each fold is used once as a test set while the remaining $K-1$ folds are used for training, yielding $K$ performance estimates that are subsequently averaged. The optimal model is selected according to the lowest validation mean squared error. For models with regularization, an additional penalty term is included in the loss function to suppress large weights, improving stability and mitigating overfitting.

\begin{table*}[t]
\renewcommand{\arraystretch}{1.3}
\caption{RMSE averaged over Pauli strings for Ridge regression and XGBoost with optimal hyper parameters obtained from Grid Search with K-Fold Cross Validation. Near-Clifford and Clifford columns correspond to models trained on the associated types of circuits, while Noisy error column indicates the error prior to mitigation.}
\vspace*{3mm}
\scalebox{1.0}{
\begin{tabular}{cccccccc}
\hline
\multicolumn{1}{|c|}{\multirow{2}{*}{}} & \multicolumn{1}{|c|}{\multirow{2}{*}{$p$}}      & \multicolumn{2}{c|}{\begin{tabular}[c]{@{}c@{}}Ridge, $\times 10^{-3}$\end{tabular}} & \multicolumn{2}{c|}{\begin{tabular}[c]{@{}c@{}}XGBoost, $\times 10^{-3}$\end{tabular}} & \multicolumn{2}{c|}{\begin{tabular}[c]{@{}c@{}}Noisy error, $\times 10^{-3}$\end{tabular}} \\ \cline{3-8} 
\multicolumn{1}{|c|}{}  & \multicolumn{1}{|c|}{}                          & \multicolumn{1}{c|}{near-Clifford}              & \multicolumn{1}{c|}{Clifford}                     & \multicolumn{1}{c|}{near-Clifford}                   & \multicolumn{1}{c|}{Clifford}                  & \multicolumn{1}{c|}{near-Clifford}                      & \multicolumn{1}{c|}{Clifford}                     \\ \hline
\multicolumn{1}{|c|}{\multirow{3}{*}{Depolarization}} & \multicolumn{1}{|c|}{0.01}                      & \multicolumn{1}{c|}{2.7}               & \multicolumn{1}{c|}{8.2}                      & \multicolumn{1}{c|}{17.9}                    & \multicolumn{1}{c|}{18.5}                   & \multicolumn{1}{c|}{15.6}                       & \multicolumn{1}{c|}{37.0}                      \\ \cline{2-8}
\multicolumn{1}{|c|}{} & \multicolumn{1}{|c|}{0.05}                      & \multicolumn{1}{c|}{11.3}               & \multicolumn{1}{c|}{43.8}                      & \multicolumn{1}{c|}{20.6}                    & \multicolumn{1}{c|}{11.3}                   & \multicolumn{1}{c|}{57.4}                       & \multicolumn{1}{c|}{124.6}                     \\ \cline{2-8}
\multicolumn{1}{|c|}{} & \multicolumn{1}{|c|}{0.1}                       & \multicolumn{1}{c|}{18.8}               & \multicolumn{1}{c|}{84.8}             & \multicolumn{1}{c|}{24.0}                    & \multicolumn{1}{c|}{16.0}                   & \multicolumn{1}{c|}{87.0}                       & \multicolumn{1}{c|}{160.9}                     \\ \hline

\multicolumn{1}{|c|}{\multirow{3}{*}{Pauli Noise}} & \multicolumn{1}{|c|}{0.01}                      & \multicolumn{1}{c|}{2.3}               & \multicolumn{1}{c|}{4.77}                      & \multicolumn{1}{c|}{18.2}                    & \multicolumn{1}{c|}{19.6}                   & \multicolumn{1}{c|}{12.6}                       & \multicolumn{1}{c|}{27.8}                      \\ \cline{2-8}
\multicolumn{1}{|c|}{} & \multicolumn{1}{|c|}{0.05}                      & \multicolumn{1}{c|}{9.5}               & \multicolumn{1}{c|}{28.3}                      & \multicolumn{1}{c|}{18.3}                    & \multicolumn{1}{c|}{17.4}                   & \multicolumn{1}{c|}{49.6}                       & \multicolumn{1}{c|}{94.7}                      \\ \cline{2-8}
\multicolumn{1}{|c|}{} & \multicolumn{1}{|c|}{0.1}                       & \multicolumn{1}{c|}{16.3}               & \multicolumn{1}{c|}{58.4}                      & \multicolumn{1}{c|}{21.6}                    & \multicolumn{1}{c|}{20.7}                   & \multicolumn{1}{c|}{73.9}                       & \multicolumn{1}{c|}{147.1}                     \\ \hline

\multicolumn{1}{|c|}{\multirow{3}{*}{Composite}} & \multicolumn{1}{|c|}{0.01}                      & \multicolumn{1}{c|}{6.8}               & \multicolumn{1}{c|}{15.1}                      & \multicolumn{1}{c|}{17.7}                    & \multicolumn{1}{c|}{52.9}                   & \multicolumn{1}{c|}{25.1}                       & \multicolumn{1}{c|}{55.3}                      \\ \cline{2-8}
\multicolumn{1}{|c|}{} & \multicolumn{1}{|c|}{0.05}                      & \multicolumn{1}{c|}{31.0}               & \multicolumn{1}{c|}{93.5}                      & \multicolumn{1}{c|}{28.4}                    & \multicolumn{1}{c|}{73.8}                   & \multicolumn{1}{c|}{79.5}                       & \multicolumn{1}{c|}{150.9}                     \\ \cline{2-8}
\multicolumn{1}{|c|}{} & \multicolumn{1}{|c|}{0.1}                       & \multicolumn{1}{c|}{50.7}               & \multicolumn{1}{c|}{113.8}             & \multicolumn{1}{c|}{43.3}                    & \multicolumn{1}{c|}{107.4}                   & \multicolumn{1}{c|}{105.7}                       & \multicolumn{1}{c|}{178.9}                     \\ \hline
\end{tabular}}
\label{crossval_rmse}
\end{table*}

In this work we are benchmarking the most popular machine learning models for the regression task: linear regression (with L1/L2 regularization~\cite{su2012linear, ranstam2018lasso, mcdonald2009ridge, hans2011elastic}, Random Forest~\cite{rigatti2017random}, SVM~\cite{jakkula2006tutorial}, KNN Regressor~\cite{song2017efficient}, MLP~\cite{taud2017multilayer} and XGBoost~\cite{chen2016xgboost}.  Under the described procedure, Linear Regression with L2-regularization (also known as Ridge regression), and XGBoost demonstrate the highest mitigation accuracy. Table \ref{crossval_rmse} depicts the results of this procedure with root mean squared errors (RMSE) averaged over Pauli string, i.e. $\sqrt{L(\phi)/\mathcal{P}}$. The results in Table~\ref{crossval_rmse} indicate that Ridge regression trained on near-Clifford data generally achieves superior error mitigation: at low error rates the effect of noise is almost eliminated, while for high noise levels its effect is substantially reduced. At higher noise levels, however, XGBoost could occasionally outperform Ridge regression. The relatively strong performance of XGBoost trained on the Clifford set may be attributed to the model's piecewise constant behavior, which aligns well with the structure of the Clifford dataset, where each $\langle P_i\rangle^{\text{ideal}} \in \{0, \pm1\}$. Despite XGBoost’s advantage in mitigating errors on Pauli strings under high noise, Ridge regression consistently exhibits notable error suppression across all tested regimes. The superior performance of Ridge regression can be attributed to its strong regularization and numerical stability. In the presence of limited training data the  penalty suppresses overfitting, leading to more robust and reliable mitigation compared to more flexible models. 

Such linear models are often accompanied by different scalers to improve their numerical stability. In our case we assist Ridge regression with a Standard Scaler, which transforms noisy Pauli strings as $\mathbf{P}^{\text{scaled}} = \mathbf{S}\mathbf{P}^{\text{noisy}} - \mathbf{c},$ where $\mathbf{S} = \text{diag}(\sigma_1^{-1}, \sigma_2^{-1}, \dotsc , \sigma_{\mathcal{P}}^{-1})$ is a diagonal matrix with $\sigma_i$ being the standard deviation of $i$-th noisy Pauli string $\langle P_i\rangle^{\text{noisy}}$, and $\mathbf{c} = (\mu_1/\sigma_1, \mu_2/\sigma_2, \dotsc, \mu_\mathcal{P}/\sigma_\mathcal{P})^T$ with $\mu_i$ being mean value of $i$-th Pauli string. All the values are computed on the training dataset. 

Thus, the final constructed map consists of linear standard scaler and consequent Ridge regression. Being a combination of two linear transforms, this map can be represented as
\begin{equation}
f_\phi(\mathbf{P}) = \mathbf{MP} + \mathbf{b},
\end{equation}
where $\mathbf{M}$ is a $\mathcal{P} \times \mathcal{P}$ matrix and $\mathbf{b}$ is a bias vector of length $\mathcal{P}$. The corresponding estimate of the Hamiltonian expectation value is then
\begin{equation}\label{energy_mitig_ridge}
\hat{E} = \mathbf{h}\cdot f_\phi(\mathbf{P}) = \mathbf{h \cdot M P} + \mathbf{h\cdot b}.
\end{equation}
After training, the off-diagonal elements of $\mathbf{M}$ are observed to be several orders of magnitude smaller than the diagonal ones. This suggests that, under the present noise model, the method does not learn significant correlations between different Pauli strings. Although the cumulative effect of the off-diagonal elements may be notable due to the number of terms, we observe no significant contribution from them.

Given the nearly-diagonal structure of $\mathbf{M}$, it is natural to ask whether the diagonal entries $M_{ii}$ exhibit substantial variation or are approximately uniform. In other words, does Ridge mitigation induce a nontrivial transformation of the cost landscape, or does it effectively reduce to a global rescaling and offset? To investigate this, we neglect the off-diagonal elements and approximate $\mathbf{M} \approx \mathrm{diag}(\mathbf{M})$. We then decompose
\begin{equation}
    \mathbf{M} = \mathbf{M}_0 + \delta\mathbf{M},
\end{equation}
where $\mathbf{M_0} = M_0 \mathbb{1}$ is a uniform diagonal matrix with $M_0 = \dfrac{1}{\mathcal{P}}\sum\limits_{i=1}^\mathcal{P}M_{ii}$ Substituting this into equation~\eqref{energy_mitig_ridge} yields
\begin{equation}
\hat{E} = M_0\mathbf{h}\cdot\mathbf{P} + \mathbf{h}\cdot\mathbf{b} + \mathbf{h}\cdot\delta\mathbf{M}\mathbf{P}.
\end{equation}
Here, the first term corresponds to a global rescaling of the noisy Hamiltonian, the second introduces a constant offset, and the third captures string-specific corrections. Numerical simulations show that this last term is comparable in magnitude to the others. That confirms that Ridge regression performs a genuinely nontrivial transformation rather than a simple global adjustment.

We also considered two modes of mitigation---mitigating the target Hamiltonian and mitigating each Pauli string independently. Experiments with Ridge regression demonstrate no significant performance difference between these two modes. At the same time, XGBoost mitigation of the Hamiltonian shows inferior performance (RMSE being severalfold larger) compared to individual Pauli string mitigation. Thus, we conclude that Pauli strings mitigation is not only more versatile in terms of covering a whole range of Hamiltonians of similar structure, but can also provide higher accuracy of mitigation.  

\subsection{ML-QEM in VQE}

The primary objective of the proposed protocol is to enhance the robustness of VQE against noise through the use of ML-QEM technique. Noise degrades VQE performance not only by reducing the accuracy of expectation value estimates but can also alter the cost-function landscape, leading to suboptimal variational parameters. This raises the question~\cite{Wang_2024} of whether (i) incorporating the mitigation procedure directly into the optimization feedback loop---mitigated optimization---provides any advantage over (ii) applying mitigation only after the optimization has converged---post-optimization mitigation.

\begin{figure}[h!]
\centering
\includegraphics[width=\linewidth]
{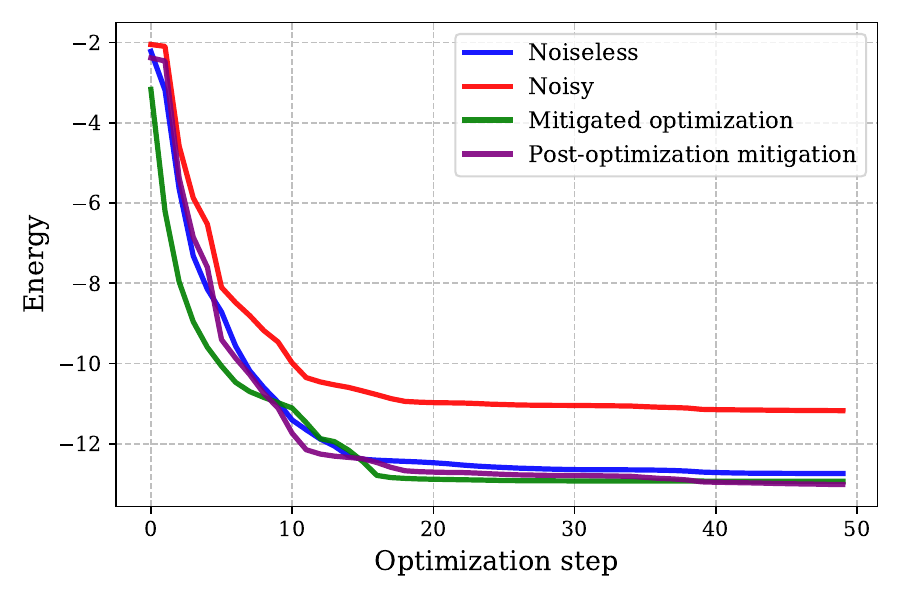}
\caption{VQE optimization dynamics for the depolarizing noise model ($p = 0.01$). Shown are noiseless (blue), noisy (red), obtained under mitigated optimization (green) and post-optimization mitigation (purple) values. The overlap of the mitigated curves demonstrates that in-loop mitigation does not improve convergence compared to post-optimization correction.}
\label{optimization}
\end{figure}

\begin{figure}[h!]
\centering
\includegraphics[width=\linewidth]
{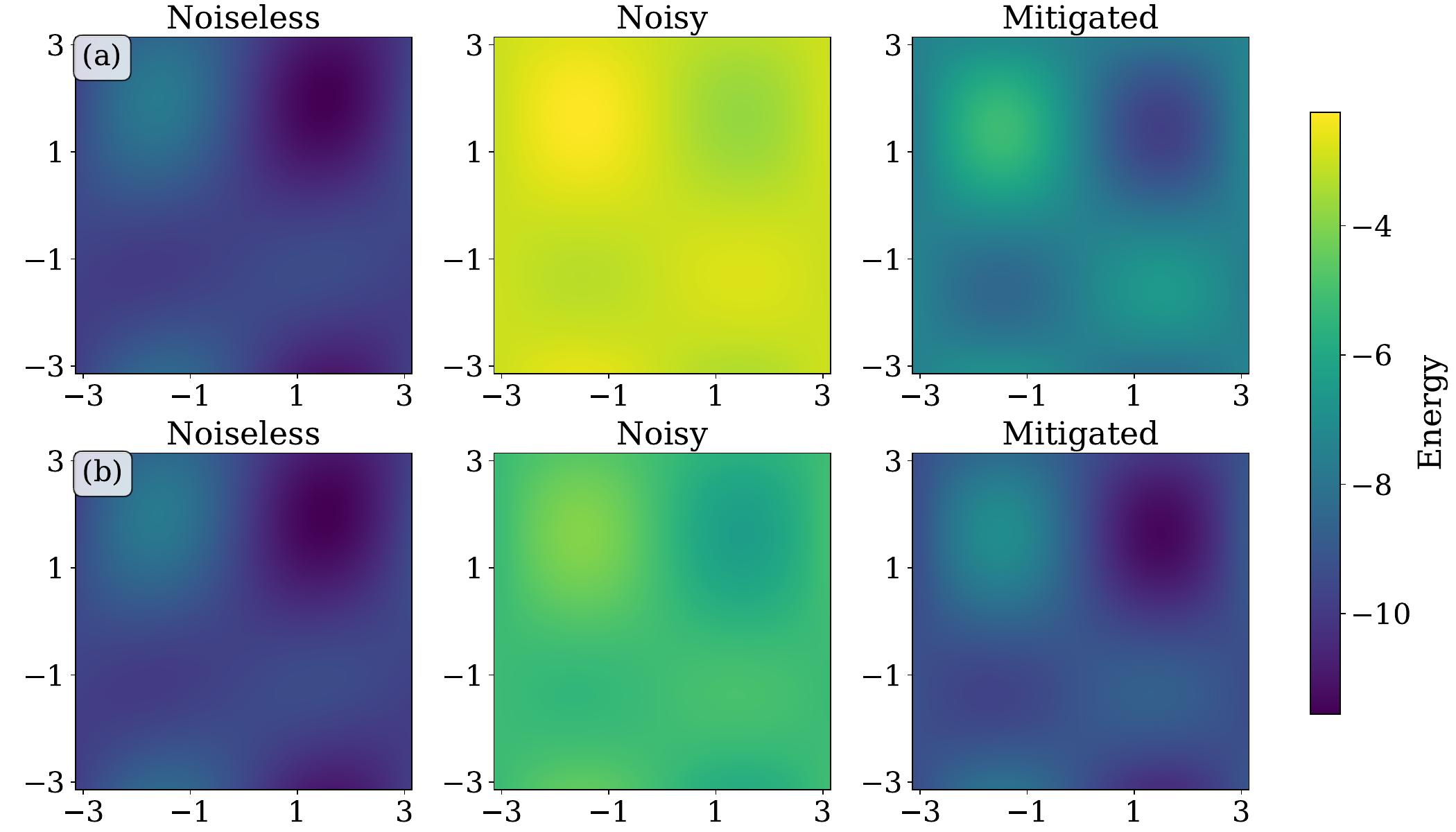}
\caption{Noiseless, noisy and mitigated cost landscapes with two random parameters varied in range $(-\pi, \pi)$ under composite noise (a) and Pauli noise (b) with strength $p=0.05$.}
\label{landscape}
\end{figure}

Due to the piecewise-constant behavior of the XGBoost mapping, which causes the optimizer to become trapped in local plateaus, only Ridge regression was considered in this study. Owing to the significant computational cost of the corresponding simulations for $n=12$ qubits, the number of performed runs was limited to 3 random Sherrington-Kirkpatrick instances for each noise model, resulting in a total of 27 runs. Nevertheless, across all trials, no observable difference was found between the two mitigation strategies: both approaches consistently converged to nearly identical energy values. This behavior was also validated across a larger statistical sample for $n=6$ qubits using 100 random instances.
The typical behavior of (i) mitigated optimization compared with (ii) the post-optimization mitigation is demonstrated in figure \ref{optimization}. Note that the post-optimization mitigation curve depicts values obtained from the ML model applied at each step of noisy mitigation, while the optimization was conducted having access only to the noisy values.  This indicates that, for the considered setting, integrating quantum error mitigation into the optimization loop does not yield a measurable improvement in performance. 
\vspace*{1mm}
To support this observation, we analyze the transformation of the VQE cost landscape under noise and subsequent mitigation. We consider a randomly generated Hamiltonian and perform noiseless VQE optimization. Two parameters are then varied over the range $(-\pi, \pi)$ to visualize  noiseless, noisy, and ML-mitigated cost function landscapes. The resulting landscapes are shown in figure \ref{landscape}.
It can be observed that the global minimum of the cost landscape is preserved under the considered noise models. While noise modifies the scale and local features of the landscape, its overall structure remains largely unchanged. This indicates that the mapping induced by noise does not significantly distort the geometry of the optimization problem in parameter space. Consequently, incorporating error mitigation directly into the optimization loop does not provide a noticeable advantage over post-optimization mitigation, as the optimizer is already guided toward the correct region of the landscape. Thus, we restrict our consideration to post-optimization mitigation as it is more computationally efficient..

\begin{figure*}[t]
\centering
\includegraphics[width=\linewidth]
{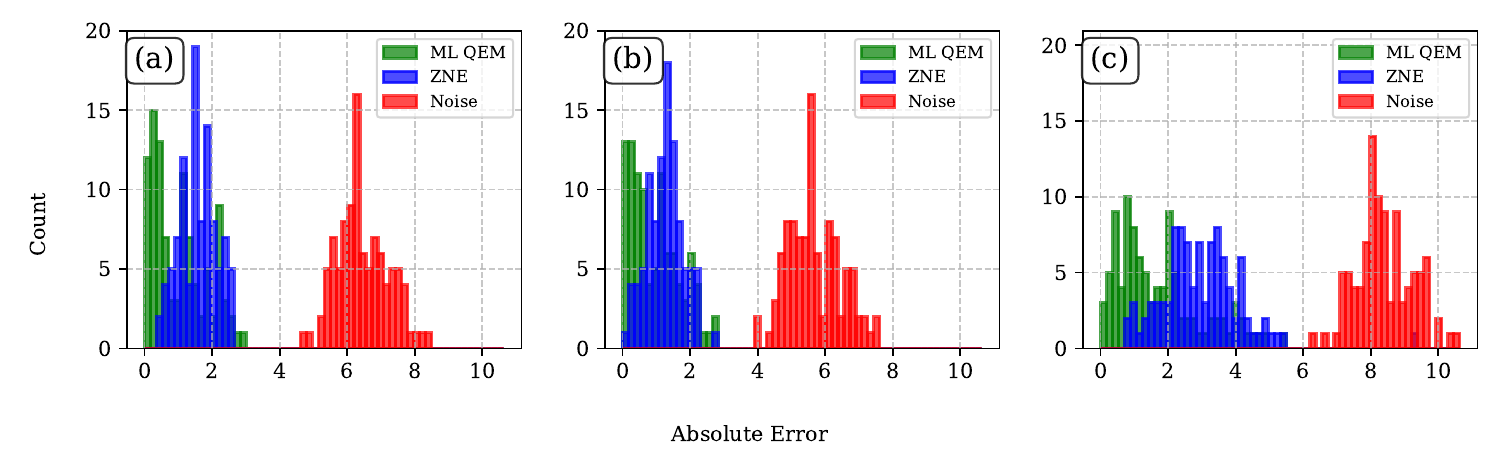}

\caption{\textbf{Absolute error $|E^{\text{ideal}} - \hat{E}|$ distribution for each considered noise model with corresponding noise strength 0.05}
Here ML-QEM depicts the results of best acquired models for each case, i.e. Ridge regression trained on near-Clifford data for depolarization (a), Pauli noise (b), and Ridge regression trained on Clifford data for the composite noise model (c).}
\label{hist}
\end{figure*}

\begin{table*}[t]
\renewcommand{\arraystretch}{1.3}
\caption{Comparison of error mitigation methods across different noise settings in terms of the error suppression factors $|E^{\text{noisy}} - E^{\text{ideal}}|/|\hat{E} - E^{\text{ideal}}|$.
Values are reported as median and [Q1--Q3] of the distributions across considered instances. Higher error suppression factors indicate better performance. Bold font indicates the best performing model.}
\label{error_suppression}
\vspace*{3mm}
\begin{tabular}{|c|c|c|c|c|c|c|}
\hline
\multicolumn{1}{|c|}{\multirow{2}{*}{$p$}}& \multicolumn{2}{c|}{Ridge} & \multicolumn{2}{|c|}{XGBoost} & \multicolumn{1}{c|}{\multirow{2}{*}{ZNE}} & \multicolumn{1}{c|}{\multirow{2}{*}{Noisy RMSE}} \\ \cline{2-5}
 & near-Clifford & Clifford & near-Clifford & Clifford &  &  \\
\hline

\multicolumn{7}{|c|}{\textbf{Depolarizing Noise}} \\ \hline
0.01  & 8.2 [5.1--14.8] & 6.6 [3.1--15.7] & 0.8 [0.6--0.9] & 1.2 [0.8--2.1] & \textbf{44.4 [36.7--56.4]} & 1.849 \\
0.05  & \textbf{8.0 [3.9--20.1]} & 3.6 [2.1--7.8] & 1.9 [1.6--2.1] & 5.0 [3.4--10.3] & 4.0 [3.5--5.1] & 6.507 \\
0.1   & 3.5 [1.8--7.5] & 2.3 [1.3--4.8] & 2.2 [1.9--2.5] & \textbf{5.3 [3.4--12.5]} & 2.6 [2.1--3.2] & 9.289 \\

\hline
\multicolumn{7}{|c|}{\textbf{Pauli Noise}} \\ \hline
0.01 & 7.8 [4.7--13.0] & 7.9 [4.6--19.9] & 0.6 [0.5--0.8] & 1.3 [0.8--3.0] & \textbf{70.0 [43.2--148.5]} & 1.588 \\
0.05 & \textbf{7.4 [3.6--14.6]} & 5.8 [3.6--12.4] & 1.9 [1.5--2.1] & 5.4 [3.0--14.7] & 4.4 [3.7--5.6] & 5.754 \\
0.1  & 4.4 [2.6--8.7] & 3.3 [2.1--7.7] & 2.0 [1.8--2.3] & \textbf{5.7 [3.5--9.7]} & 2.8 [2.3--3.5] & 8.439 \\



\hline
\multicolumn{7}{|c|}{\textbf{Composite}} \\ \hline
0.01 &7.8 [4.5-14.0] & 5.9 [2.7-13.3] & 1.1 [0.9-1.3] & 0.9 [0.7-1.4] & \textbf{19.0 [15.1-26.4]} & 2.886 \\
0.05 & 5.5 [3.3-9.8] & 4.2 [2.8-9.4] & 1.8 [1.7-2.1] & \textbf{6.1 [3.4-10.6]} & 2.8 [2.4-3.6] & 8.446 \\
0.1  & \textbf{3.4 [2.2-6.3]} & 2.5 [1.9-3.7] & 1.6 [1.4-1.7] & 2.4 [1.7-3.0] & 1.0 [1.0-1.0] & 10.825 \\

\hline
\end{tabular}
\end{table*}

In order to benchmark the proposed method, $K = 100$ instances of Sherrington-Kirkpatrick Hamiltonians \cite{panchenko2012sherrington, sherrington1975solvable} are generated by sampling $J_{ij} \sim \mathcal{N}(0,1)$. For each Hamiltonian, first the noiseless simulation of VQE optimization is run with \texttt{TwoLocal} ansatz, the optimization is done with the L-BFGS-B optimizer \cite{liu1989limited}. Then the parameters obtained from the optimization are used to execute VQE circuit subjected to different noise models. The resultant expectations are then mitigated using the trained ML-QEM model. Figure ~\ref{hist} illustrates typical energy error distributions obtained from noisy simulations and after the mitigation. Error suppression by about an order of magnitude can be clearly seen across all considered noise models. For a more informative characterization, in table~\ref{error_suppression} we present statistics of error suppression factors $|E^{\text{noisy}} - E^{\text{ideal}}|/|\hat{E} - E^{\text{ideal}}|$, which characterize how errors are suppressed across the considered random problem Hamiltonians. The distribution of error suppression factors is an informative metric for evaluating the efficiency of quantum error mitigation methods, as it directly quantifies error reduction across instances. A fraction of instances where the error suppression factor falls below 1 (i.e.~the cases where error mitigation actually worsened the performance) also characterizes the probability of protocol failure. In Appendix~\ref{rmse_vqe} we also present the same results in the form of RMSE.

The results demonstrate that the proposed protocol achieves significant error reduction in the low-noise regime. As the noise strength increases, the mitigation remains robust, consistently yielding a severalfold reduction across all considered settings. Ridge regression trained on near-Clifford data continues to deliver strong performance in most noise regimes, highlighting the effectiveness of training on datasets that provide broad coverage of the relevant region of Hilbert space. In the high-noise regime, its performance is surpassed by XGBoost in the case of depolarizing noise and Pauli noise. This may be attributed to the increasing contribution of nonlinear noise effects to the VQE output, which are more effectively captured and mitigated by a nonlinear model. Nevertheless, even in these more challenging settings, the proposed method maintains stable and notable suppression of error.

To benchmark the obtained results, we compare the proposed protocol with the standard unitary folding ZNE. We find that it outperforms the proposed method in the low-noise regime. However, as the noise strength increases, the advantage of ZNE diminishes, while the proposed method remains stable, ultimately becoming comparable  and superior to ZNE. These results highlight the robustness and competitiveness of the proposed approach, particularly under increasing noise. Nevertheless, it is worth noting that the performance comparison of ML-QEM and ZNE can depend on a particular noise model considered. As such, under the amplitude damping ($T_1$) noise, ML-QEM demonstrates error suppression comparable to the composite model, but becomes inferior to ZNE.  Implementation details of ZNE are provided in Appendix \ref{zne}.

We analyze how the probabilistic generation of near-Clifford circuits—specifically, inserting Haar unitaries with a probability $q$—affects mitigation accuracy. For both depolarizing and composite noise models with a noise strength of $p = 0.05$, we execute the entire protocol across a range of $q \in \{0.2, 0.4, 0.6, 0.8\}$, which includes dataset generation, hyperparameter optimization, and VQE benchmarking. The results demonstrate that while XGBoost mitigation accuracy is superior for pure Clifford circuits, Ridge regression performance does not depend significantly on the fraction of non-Clifford gates. Consequently, the proposed approach scales efficiently with the number of qubits.

The proposed scheme for dataset generation would actually yield the same data for different ansatze with the same entangling layer nearest neighbor topology. Thus, the resulting ML-QEM models are also applicable to a broad range of ansatze. To test this statement, the same protocols with previously generated 100 Sherrington-Kirkpatrick Hamiltonians is run for the \texttt{TwoLocal} ansatz with each $R_Y$ rotating gate being replaced as $R_Y \to R_ZR_YR_Z$ giving rise to a general single-qubit rotation. The pipeline is analogous: for each Hamiltonian the VQE ansatz is optimized on the noiseless simulator, then the obtained circuit is run on the noisy simulator and finally the measured noisy Pauli strings are summed into the Hamiltonian of consideration. Results demonstrate the same performance as in the case of $R_Y$ rotating layer demonstrating the versatility of the proposed model which, once trained, can be applied to a broad range of ansatze and Hamiltonians.
\vspace*{1mm}
\section{Conclusion}
\label{conclusion}

In this work, we proposed and benchmarked a practically oriented ML-QEM protocol for VQE. Training datasets were generated using Clifford and near-Clifford circuits, evaluated across various ML models via a grid search with K-fold cross-validation. Our results reveal a crucial interplay between dataset characteristics and model performance. Ridge regression achieved peak accuracy when trained on near-Clifford data, as the continuous Hilbert space coverage of Haar unitaries provides a highly informative dataset for capturing complex gate distortions. Crucially, by introducing these unitaries probabilistically, our protocol could maintain classical simulation scalability while preserving the mitigation accuracy of the trained model. Conversely, XGBoost demonstrated unique resilience when trained on pure Clifford sets, aligning well with the discrete, piecewise-constant nature of Clifford target values. Overall, Ridge regression and XGBoost performed the best, with Ridge regression consistently leading in most cases.

We considered two approaches to ML-based error mitigation: applying mitigation directly to the target Hamiltonian and applying it to individual Pauli string expectation values. The results showed that, depending on the choice of ML model, mitigating individual Pauli strings performed comparably to—or better than—target Hamiltonian mitigation. This indicates that Pauli string mitigation not only provides greater flexibility---once trained, the model can be applied to a broad family of Hamiltonians---but also achieves superior mitigation accuracy. 
Under the considered noise models, the resulting linear map was found to be nearly diagonal, with off-diagonal elements providing minor contribution to the final results. Nevertheless, retaining the full matrix form yielded improved accuracy and was therefore preferable, while remaining computationally efficient.

Finally, we benchmarked the proposed method in the VQE setting using 100 instances of $n=12$ qubit Sherrington–Kirkpatrick Hamiltonians with transverse field across various noise configurations and compared it with standard unitary folding–based ZNE. The protocol demonstrated a several-fold reduction in error, outperforming ZNE as circuit noise increased. 
Moreover, ZNE relies on evaluating multiple noise-scaled circuits, e.g.~via unitary folding, which increases circuit depth and results in longer effective execution times. This, in turn, imposes stricter requirements on qubit coherence times, increases sampling overhead due to repeated circuit evaluations, and demands precise experimental control to ensure reliable noise scaling, making its practical deployment on current NISQ devices more challenging. ML-QEM, on the contrary, does not require precise control over the noise in the system and does not increase circuit depth, but requires executing many similar circuits to collect the training data. However, once trained, the model applies to a wide range of alike problem Hamiltonians (composed of the considered Pauli strings) and variational ansatze of similar structure, whereas ZNE requires running for every combination of circuit and observable.

{\color{red}}

We also observed that (i) applying mitigation at every optimization step and (ii) applying mitigation after the optimization provided largely similar results. This conclusion was further supported by the analysis of how circuit noise and subsequent mitigation affect the cost function landscape. Specifically, we observed that, while the scale and local features of the landscape were modified under the considered noise models, the global minimum remained unchanged. Thus, incorporating mitigation directly into the optimization loop did not alter the optimal circuit parameters, leading to similar performance. However, for more complex noise models, where the cost function landscape may be distorted more substantially,
the placement of the mitigator within the feedback loop may become more critical. In such cases, mitigation applied during optimization could influence the optimization trajectory itself and lead to different convergence behavior and final solutions. Using ZNE in such scenario would require performing unitary folding per each set of parameters appearing during the optimization, which negatively affects compatibility of ZNE with variational computing.


A key challenge in applying ML-based QEM lies in constructing the training dataset, as noiseless expectation values of observables cannot be directly obtained from quantum hardware. In this work, we focused on applying ML-QEM to VQE, which imposes an additional challenge: the trained model must generalize across parametrized circuits for arbitrary parameter values. In the work \cite{czarnik2021} authors propose using near-Clifford data for training ML-QEM models. However, authors focus on mitigating fixed circuit which provides less versatility for variational computing and also do not consider other ML algorithms. Another work \cite{liao2024} provides extensive tests of different models and considers the application of the proposed method to variational algorithms. However, this consideration is limited to a small system size and lacks tests across different levels of noise and does not go into detail on the practical aspects of training dataset gathering. The work \cite{babukhin2024echo} proposes a so-called echo-evolution method for generating the dataset for training neural networks to mitigate quantum evolution. However, only neural networks were considered in this work, overlooking other algorithms, and the method is not transferable to variational computing.

In our work this issue was addressed with a practical dataset generation protocol based on near-Clifford circuits, enabling the training of models that generalize across the full parameter space and are not restricted to a specific ansatz or Hamiltonian. Overall, our results demonstrate that simple, data-driven approaches—particularly well-regularized linear models—provide robust, practical, and versatile error mitigation for variational computing in realistic NISQ settings.

\section{Data and Code availability}
The data and the source code to reproduce all numerical simulations are available in a public GitHub repository: \url{https://github.com/boochmo/ml-qem-clifford}.

\section*{Acknowledgements}
We thank Zakhar Sayapin for stimulating discussions. 

\bibliography{refs.bib}
\bibliographystyle{unsrt}

%
\appendix
\section{RMSE for ML-QEM benchmark in VQE}
\label{rmse_vqe}
The results of benchmarking the proposed protocol under different noise models and comparing it to ZNE, given in the table~\ref{error_suppression} are represented here in the RMSE format, i.e. 
\begin{equation}
\sqrt{\dfrac{1}{N}\sum\limits_{i}^N\left(\hat{E}_i - E^\text{ideal}\right)},    
\end{equation}
in the table~\ref{rmse}. Despite the different format, it induces the same conclusions: despite stronger ZNE error reduction in the lower noise regime, the proposed method shows superior results and notable error suppression upon noise strength increase.
\begin{table}[H]
\centering
\renewcommand{\arraystretch}{1.3}
\caption{Comparison of Root Mean Squared Errors (RMSE) of trained models between combat, initial noise error and ZNE at 100 Hamiltonians of the Sherrington-Kirkpatrick. Bold number represent the best results}
\vspace{5pt}
\scalebox{0.85}{
\begin{tabular}{ccccccc}
\hline
\multicolumn{7}{|c|}{Depolarizing Noise}                                                                                                                                                                                                                                                                                                  \\ \hline
\multicolumn{1}{|c|}{\multirow{2}{*}{$p$}} & \multicolumn{2}{c|}{Ridge}                                                & \multicolumn{2}{c|}{XGBoost}                                      & \multicolumn{1}{c|}{\multirow{2}{*}{ZNE}} & \multicolumn{1}{c|}{\multirow{2}{*}{\begin{tabular}[c]{@{}c@{}}Noisy\\ error\end{tabular}}} \\ \cline{2-5}
\multicolumn{1}{|c|}{}                          & \multicolumn{1}{c|}{near-Clifford}           & \multicolumn{1}{c|}{Clifford}           & \multicolumn{1}{c|}{near-Clifford}   & \multicolumn{1}{c|}{Clifford}           & \multicolumn{1}{c|}{}                                                                       & \multicolumn{1}{c|}{}                     \\ \hline
\multicolumn{1}{|c|}{0.01}                      & \multicolumn{1}{c|}{0.308} & \multicolumn{1}{c|}{0.41}          & \multicolumn{1}{c|}{2.433}  & \multicolumn{1}{c|}{1.944}          & \multicolumn{1}{c|}{0.045}                & \multicolumn{1}{c|}{1.849}                                                                  \\ \hline
\multicolumn{1}{|c|}{0.05}                      & \multicolumn{1}{c|}{1.296} & \multicolumn{1}{c|}{2.392}          & \multicolumn{1}{c|}{3.547} & \multicolumn{1}{c|}{1.7671}          & \multicolumn{1}{c|}{1.668}               & \multicolumn{1}{c|}{6.507}                                                                 \\ \hline
\multicolumn{1}{|c|}{0.1}                       & \multicolumn{1}{c|}{4.553}         & \multicolumn{1}{c|}{7.061}         & \multicolumn{1}{c|}{4.293} & \multicolumn{1}{c|}{2.510} & \multicolumn{1}{c|}{5.371}               & \multicolumn{1}{c|}{9.289}                                                                 \\ \hline
\multicolumn{1}{l}{}                            & \multicolumn{1}{l}{}                & \multicolumn{1}{l}{}                & \multicolumn{1}{l}{}        & \multicolumn{1}{l}{}                & \multicolumn{1}{l}{}                      & \multicolumn{1}{l}{}                                                                        \\ \hline

\multicolumn{7}{|c|}{Pauli Noise}                                                                                                                                                                                                                                                                                                         \\ \hline
\multicolumn{1}{|c|}{\multirow{2}{*}{$p$}}        & \multicolumn{2}{c|}{Ridge}                                                & \multicolumn{2}{c|}{XGBoost}                                      & \multicolumn{1}{c|}{\multirow{2}{*}{ZNE}} & \multicolumn{1}{c|}{\multirow{2}{*}{\begin{tabular}[c]{@{}c@{}}Noisy\\ error\end{tabular}}} \\ \cline{2-5}
\multicolumn{1}{|c|}{}                          & \multicolumn{1}{c|}{near-Clifford}           & \multicolumn{1}{c|}{Clifford}           & \multicolumn{1}{c|}{near-Clifford}   & \multicolumn{1}{c|}{Clifford}           & \multicolumn{1}{c|}{}                                                                       & \multicolumn{1}{c|}{}                     \\ \hline
\multicolumn{1}{|c|}{0.01}                      & \multicolumn{1}{c|}{0.297} & \multicolumn{1}{c|}{0.275}          & \multicolumn{1}{c|}{2.648}  & \multicolumn{1}{c|}{1.649}          & \multicolumn{1}{c|}{0.031}                & \multicolumn{1}{c|}{1.588}                                                                  \\ \hline
\multicolumn{1}{|c|}{0.05}                      & \multicolumn{1}{c|}{1.248} & \multicolumn{1}{c|}{1.409}          & \multicolumn{1}{c|}{3.245} & \multicolumn{1}{c|}{1.546}          & \multicolumn{1}{c|}{1.394}               & \multicolumn{1}{c|}{5.754}                                                                 \\ \hline
\multicolumn{1}{|c|}{0.1}                       & \multicolumn{1}{c|}{2.999} & \multicolumn{1}{c|}{3.180}         & \multicolumn{1}{c|}{4.235} & \multicolumn{1}{c|}{2.170}          & \multicolumn{1}{c|}{3.379}               & \multicolumn{1}{c|}{8.439}                                                                 \\ \hline
                                                                  \\ \hline
\multicolumn{7}{|c|}{Composite}                                                                                                                                                                                                                                                                                                    \\ \hline
\multicolumn{1}{|c|}{\multirow{2}{*}{$\gamma$}}        & \multicolumn{2}{c|}{Ridge}                                                & \multicolumn{2}{c|}{XGBoost}                                      & \multicolumn{1}{c|}{\multirow{2}{*}{ZNE}} & \multicolumn{1}{c|}{\multirow{2}{*}{\begin{tabular}[c]{@{}c@{}}Noisy\\ error\end{tabular}}} \\ \cline{2-5}
\multicolumn{1}{|c|}{}                          & \multicolumn{1}{c|}{near-Clifford}           & \multicolumn{1}{c|}{Clifford}           & \multicolumn{1}{c|}{near-Clifford}   & \multicolumn{1}{c|}{Clifford}           & \multicolumn{1}{c|}{}                                                                       & \multicolumn{1}{c|}{}                     \\ \hline
\multicolumn{1}{|c|}{0.01}                      & \multicolumn{1}{c|}{0.525}          & \multicolumn{1}{c|}{0.727}          & \multicolumn{1}{c|}{2.844}  & \multicolumn{1}{c|}{3.534}         & \multicolumn{1}{c|}{ 0.173}       & \multicolumn{1}{c|}{2.886}                                                                  \\ \hline
\multicolumn{1}{|c|}{0.05}                      & \multicolumn{1}{c|}{2.312}          & \multicolumn{1}{c|}{2.747} & \multicolumn{1}{c|}{4.589}  & \multicolumn{1}{c|}{2.155}          & \multicolumn{1}{c|}{3.230}                & \multicolumn{1}{c|}{8.447}                                                                 \\ \hline
\multicolumn{1}{|c|}{0.1}                       & \multicolumn{1}{c|}{4.215}         & \multicolumn{1}{c|}{5.035} & \multicolumn{1}{c|}{7.041} & \multicolumn{1}{c|}{5.332}         & \multicolumn{1}{c|}{10.591}               & \multicolumn{1}{c|}{10.826}                                                                 \\ \hline
\multicolumn{1}{l}{}                            & \multicolumn{1}{l}{}                & \multicolumn{1}{l}{}                & \multicolumn{1}{l}{}        & \multicolumn{1}{l}{}                & \multicolumn{1}{l}{}                      & \multicolumn{1}{l}{}     
\end{tabular}}
\label{rmse}
\end{table}
\section{Zero Noise Extrapolation}
\label{zne}
For comparison with the proposed method, ZNE is implemented for the considered quantum circuits. ZNE is a widely used error mitigation technique that involves evaluating the same quantum circuit at artificially amplified noise levels and extrapolating the observable of interest back to the zero-noise limit.

ZNE relies on the assumption that the effect of noise on an observable $X$ can be modeled as a function of a noise parameter $\lambda$, where $\lambda=0$ corresponds to the noiseless case. Under certain conditions—such as time-independent noise channels—the effective noise strength can be scaled by stretching the circuit in time or, more practically, by inserting identity operations or folding circuit layers. Let $X(\lambda)$ denote the expectation value of $X$ at noise level $\lambda$. The objective is to estimate $X(0)$ using measurements obtained at $\lambda > 0$. A common approach is polynomial extrapolation, where one assumes $X(\lambda)$ can be approximated by a low-degree polynomial in $\lambda$. In our implementation, we use exponential extrapolation.

Gate folding is employed to artificially increase the noise level. Given a circuit composed of gates $\{G_i\}_{i=1}^{L}$, a folded version is constructed by replacing each gate according to $G_i \to G_i(G_i^\dagger G_i)^k$, resulting in scaling the noise from its base value $\lambda = 1$ to $\lambda = 1 +2k$. In our work, we scale the noise in the range $\lambda \in \{1,3,5\}$.
\end{document}